\documentclass[prx,aps,twocolumn,reprint,amsmath,floatfix,amssymb,showpacs,superscriptaddress]{revtex4-1}

\usepackage{amsmath}
\usepackage{booktabs}
\usepackage{array}

\usepackage{ulem}
\usepackage{microtype}
\usepackage{wasysym}
\usepackage[varg]{txfonts}
\usepackage[percent]{overpic}
\usepackage{mathrsfs}
\usepackage{braket}
\usepackage{titlesec}
\usepackage{bm}
\usepackage{comment}
\usepackage{verbatim}

\usepackage{array}
\newcommand{\PreserveBackslash}[1]{\let\temp=\\#1\let\\=\temp}
\newcolumntype{C}[1]{>{\PreserveBackslash\centering}p{#1}}
\newcolumntype{R}[1]{>{\PreserveBackslash\raggedleft}p{#1}}
\newcolumntype{L}[1]{>{\PreserveBackslash\raggedright}p{#1}}

\usepackage{color}
\usepackage[colorlinks,bookmarks=false,citecolor=blue,linkcolor=blue,urlcolor=blue,pdfstartview=FitH]{hyperref}

\definecolor{darkred}{rgb}{0.7,0.0,0.0}

\definecolor{darkblue}{rgb}{0,0.02,0.45}

\definecolor{darkgreen}{rgb}{0.02,0.45,0.0}

\definecolor{violet}{rgb}{0.8,0.2,0.6}

\usepackage{amsmath}
\usepackage{amssymb}
\usepackage{graphicx}
\usepackage{subfigure}
\usepackage{pict2e}
\usepackage{multirow}
\usepackage{soul}

\newcommand{\be}{\begin{equation}}
\newcommand{\ee}{\end{equation}}
\newcommand{\bea}{\begin{eqnarray}}
\newcommand{\eea}{\end{eqnarray}}
\newcommand{\sbe}{\small\begin{equation}}
\newcommand{\see}{\end{equation}\normalsize}
\newcommand{\sbea}{\small\begin{eqnarray}}
\newcommand{\seea}{\end{eqnarray}\normalsize}
\graphicspath{{Figures/}}

\graphicspath{{Figures/}}

\def\vec{\mathbf}
\def\mc{\mathcal}
\def\nn{\nonumber}
\begin{document}
\setstcolor{red}

\title{Reentrant incommensurate order and anomalous magnetic torque in the Kitaev magnet $\beta$-$\text{Li}_2\text{IrO}_3$}

\author{Mengqun Li}
\affiliation{School of Physics and Astronomy, University of Minnesota, Minneapolis, Minnesota 55455, USA}

\author{Ioannis Rousochatzakis}
\affiliation{Department of Physics, Loughborough University, Loughborough LE11 3TU, United Kingdom}

\author{Natalia B. Perkins}
\affiliation{School of Physics and Astronomy, University of Minnesota, Minneapolis, Minnesota 55455, USA}
\date{\today}

\begin{abstract}    
We present a theoretical study of the response of $\beta$-$\text{Li}_2\text{IrO}_3$ under external magnetic fields in the $ab$, $bc$ and $ac$ crystallographic planes. The results are based on the minimal nearest-neighbor $J$-$K$-$\Gamma$ model and reveal a rich intertwining of field-induced phases and  magnetic phase transitions with distinctive signatures that can be probed directly via torque magnetometry. 
Most saliently, we observe: i) an unusual reentrance of the incommensurate counter-rotating order for fields in the $ab$-plane, and ii) a set of abrupt torque discontinuities which are particularly large for fields rotating in the $bc$ plane, and whose characteristic shape resembles closely the ones observed in the 3D Kitaev magnet $\gamma$-$\text{Li}_2\text{IrO}_3$.
An experimental confirmation of these predictions will pave the way for an accurate determination of all relevant microscopic parameters of this 3D Kitaev magnet.
\end{abstract}

\maketitle

\vspace*{-1cm}
\section{Introduction}\label{sec:Intro}
\vspace*{-0.3cm}
In recent years, a significant experimental and theoretical effort has been devoted to realization of the Kitaev quantum spin liquid phase~\cite{Kitaev2006} in 4$d$ and 5$d$ magnetic insulators~\cite{Jackeli2009,Jackeli2010,BookCao,Rau2016,Trebst2017,Knolle2017,Winter2017,Takagi2019,Motome2020a,Motome2020b}. In these materials, the interplay between the strong spin-orbit coupling and electronic correlations    gives rise to highly anisotropic and spatially dependent interactions between effective moments $J_{\text{eff}}\!=\!1/2$. In  most of the relevant, two- and three-dimensional tri-coordinated lattice structures with edge-sharing octahedra, the dominant exchange interaction is stabilized through the mechanism elucidated by  Jackeli and Khaliullin~\cite{Jackeli2009} and has a form of the bond-dependent Ising-like Kitaev interaction, which makes these materials potential candidates for exploration of  the Kitaev quantum spin liquid. However, most of these materials exhibit  complex  long-range magnetic orders at sufficiently low temperatures, indicating that other subdominant interactions between magnetic moments are present and may also have non-trivial bond-dependent character. Both experiment ~\cite{Modic2014,Modic2017,Leahy2017,Banerjee2017,Baek2017,Zheng2017,Wolter2017,Sears2017,Kasahara2018,Modic2018a,Das2019,
Majumder2019,Majumder2019b,Ruiz2017,Ruiz2019} and theory~\cite{Janssen2016,Chern2017,Janssen2019,Rousochatzakis2018,Li2019,Riedl2019,HYLee2019} have shown that  these orders are fragile  and can be efficiently suppressed by external magnetic field. It was also found that  the competition between the external field and anisotropic bond-dependent exchange interactions gives rise to  highly-anisotropic magnetization processes and stabilizes a variety of complex orders at intermediate fields~\cite{Janssen2016,Chern2017,Janssen2019,Rousochatzakis2018,Li2019,Riedl2019,HYLee2019}.

Among various experimental techniques, the torque magnetometry is especially promising tool for studying field-induced phases~\cite{Leahy2017,Modic2018a,Xing2019,Das2019,Ruiz2017,Ruiz2019,Modic2014,Modic2017}. Recently, torque magnetometry has been used in studying the field response of  several Kitaev materials, including $\alpha$-$\text{RuCl}_3$~\cite{Leahy2017,Modic2018a}, $\text{Na}_2\text{IrO}_3$~\cite{Das2019}, $\beta$-$\text{Li}_2\text{IrO}_3$~\cite{Ruiz2019} and $\gamma$-$\text{Li}_2\text{IrO}_3$~\cite{Modic2017,Modic2014}. Rather generally, the measured torque in these materials depends strongly on the magnitude of the field, exhibits anomalous behavior at critical  fields associated with field-induced transitions, and shows a sawtooth-like angular dependence at high fields. 
 
While the angular and field dependences of the torque in $\alpha$-$\text{RuCl}_3$ has been studied in Ref.~\cite{Riedl2019}, the origin of anomalous torque response in the 3D iridates remains unclear. Therefore, the  purpose of this paper is to understand the recent torque measurements in $\beta$-$\text{Li}_2\text{IrO}_3$~\cite{Ruiz2019} and $\gamma$-$\text{Li}_2\text{IrO}_3$~\cite{Modic2014,Modic2017}. These two materials appear to have closely related local energetics~\cite{Kruger2019}, and we will therefore focus entirely on the hyperhoneycomb $\beta$-Li$_2$IrO$_3$~\cite{Biffin2014a,Takayama2015,Ruiz2017,Ruiz2019, Majumder2019,Majumder2019b}, whose microscopic minimal model, the nearest-neighbor  (NN) $J$-$K$-$\Gamma$ model, has been better understood~\cite{Lee2015,Lee2016,Ducatman2018,Rousochatzakis2018,Li2019}.

\begin{figure}[!b]
{\includegraphics[width=\linewidth]{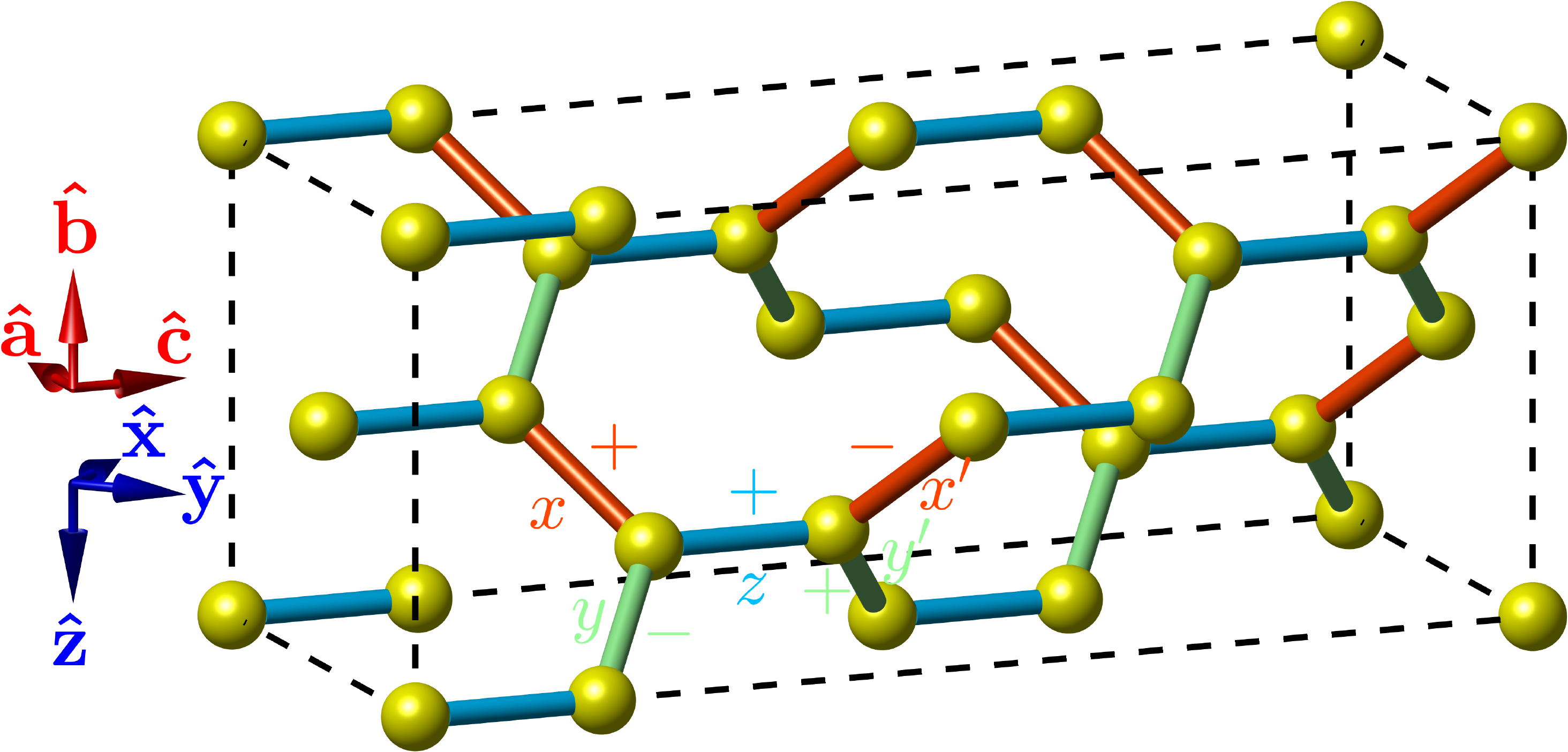}}
\caption{Orthorhombic unit cell of the hyperhoneycomb lattice. The five NN bonds of the $J$-$K$-$\Gamma$ model are marked in red for $t\in\{x,x'\}$, green for $t\in\{y,y'\}$, and blue for $t\in\{z\}$.}
\label{fig:lattice}
\end{figure}

\begin{figure*}[!t]
{\includegraphics[width=\textwidth]{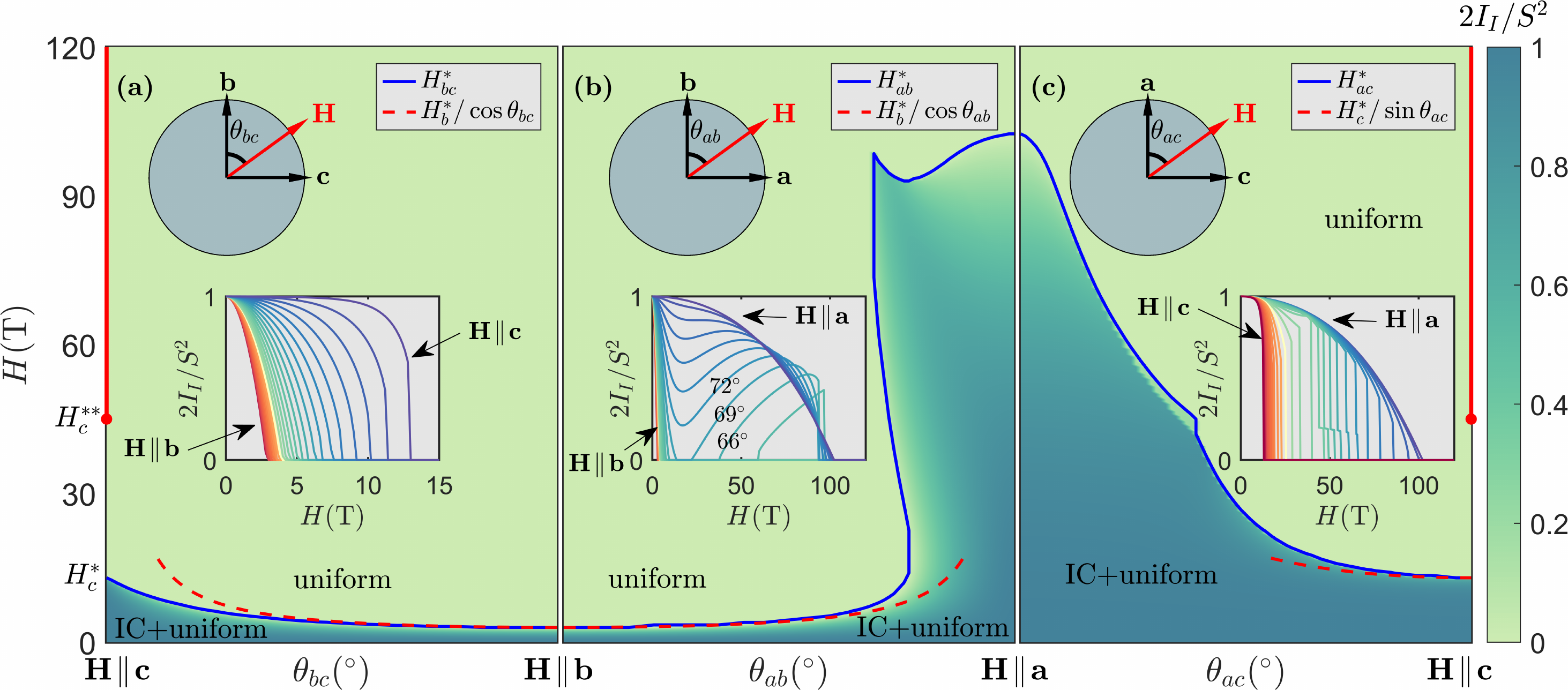}}
\caption{Global phase diagram as a function of magnetic field strength $H$ (vertical axes) and angular orientation (horizontal axes) as we rotate in the $bc$ plane (a), $ab$ plane (b), and $ac$ plane (c). The color coding tracks the evolution of the Bragg peak intensity $I_I$ of the modulated IC order, with the thick blue solid line showing its boundary, i.e., the critical field $H^*$. The red vertical line at ${\bf H}\!\parallel\!{\bf c}$ depicts the fully polarized state at $H\!\geqslant\!H_c^{**}$. The insets show the field dependence of $I_I$ at some representative angles.}\label{fig:PhaseDiagram}
\end{figure*}

To understand the behavior of the torque one first needs to elucidate the ground state phase diagram for different field orientations and strengths. To this end, we generalize the semi-analytical framework developed previously in \cite{Ducatman2018,Rousochatzakis2018,Li2019} to study the field-induced phase diagram for fields in the $ab$, $bc$ and $ac$ crystallographic planes. 
This framework builds on the idea that the incommensurate  (IC) counter-rotating order observed experimentally at zero field can be treated as a long-distance twisting of a nearby commensurate order with six spin sublattices~\cite{Ducatman2018}. 
This approach has already been applied successfully for fields along the orthorhombic directions ${\bf a}$, ${\bf b}$ and ${\bf c}$~\cite{Rousochatzakis2018,Li2019}. It has been shown in particular that the three phase diagrams share a number of qualitative features, including: i) a strong intertwining of the modulated, counter-rotating order with a set of uniform orders, some of which give rise to a finite torque signal, ii) the disappearance of the modulated order at a critical field $H^\ast$, whose value is strongly anisotropic with $H_{\bf b}^\ast\!<\!H_{\bf c}^\ast\!\ll\!H_{\bf a}^\ast$, and iii) the presence of a robust zigzag phase above $H^\ast$. It was further shown that, for ${\bf H}\!\parallel\!{\bf c}$, the disappearance of the modulated order proceeds via a first-order transition which does not restore all broken symmetries, which in turn gives rise to a second transition at $H_c^{**}$ that survives at finite temperatures~\cite{Li2019}.

The extension of this approach for fields applied in the $ab$, $bc$ and $ac$ planes shows that the interpolation of the phase diagram from one crystallographic axis to another comes with a number of unexpected results, including: i) a reentrant IC phase that survives in a finite region of field strengths and orientations for fields in the $ab$ plane (and to a lesser extent in the $ac$ plane), ii) a region featuring a two-step disappearance of the IC order, proceeding via two consecutive metamagnetic jumps, and iii) large discontinuous reversals of the magnetic torque, with characteristic sawtooth-like angular profiles that resemble closely the ones reported for $\gamma$-Li$_2$IrO$_3$. 
These results will be elucidated in Secs.~\ref{sec:PhaseDiagram} and \ref{sec:torque_abc}, but we first recap the main aspects of the microscopic model in Sec.~\ref{sec:Model}.

\vspace*{-0.3cm}
\section{Microscopic model}\label{sec:Model}
\vspace*{-0.3cm}
The Hamiltonian of the $J$-$K$-$\Gamma$ model reads~\cite{Lee2015,Lee2016,Ducatman2018,Rousochatzakis2018,Li2019}
\bea\label{eq:Hamiltonian}
\renewcommand*{\arraystretch}{1.35}
\begin{array}{c}
\mc{H}\!=\!\sum_t\sum_{\langle ij\rangle\in t}\mc{H}_{ij}^t+\mc{H}^\text{Z}\,,\\
\mc{H}_{ij}^t=J\vec{S}_i\cdot\vec{S}_j+K S_i^{\alpha_t}S_j^{\alpha_t}+\sigma_t\Gamma(S_i^{\beta_t}S_j^{\gamma_t}+S_i^{\gamma_t}S_j^{\beta_t})\,,\\
\mc{H}^\text{Z}=-\mu_B\vec{H}\cdot\sum_{i}\vec{g}_i\cdot \vec{S}_i\,.
\end{array}
\eea
Here $t \in\{x,x',y,y',z\}$ labels the different types of NN bonds (see Fig.~\ref{fig:lattice}), $\mc{H}_{ij}^t$ includes the couplings between NN sites $i$ and $j$ of a type-$t$ bond, $\alpha_t$ is the Cartesian component $x$, $y$, or $z$ associated with $t$, and $(\beta_t,\gamma_t)$ label the remaining two Cartesian components coupled by the $\Gamma$ interaction. The prefactor $\sigma_t$ alternates between $+1$ and $-1$ for $t\in\{x,y',z\}$ and $t\in\{y,x'\}$, respectively. The term $\mc{H}^\text{Z}$ stands for the Zeeman interaction, where the ${\bf g}$-tensors carry a staggered, site-dependent off-diagonal element $g_{ab}$~\cite{Ruiz2017} (in the orthorhombic frame), 
\begin{align}\label{eq:gtensor}
\vec{g}_{i}={\bf g}_{\rm diag} + p_i {\bf g}_{\rm off-diag} \equiv 
\begin{pmatrix}
	g_{aa} & p_i g_{ab} & 0\\
	p_i g_{ab} & g_{bb} & 0\\
	0 & 0 & g_{cc}\\
\end{pmatrix}
\end{align}
where $p_i\!=\!1$ for spins on the $xy$ chains and $-1$ for spins on the $x'y'$ chains (see Fig.~\ref{fig:lattice}). More specifically, for fields in the $bc$, $ab$ and $ac$ planes, the Zeeman terms read (with $h\!=\!\mu_B H$) 
\small
\bea
\!\!&&\!\!\mc{H}_{bc}^\text{Z}\!=\!-h\!\!\sum_i \!\Big[\!
\cos \theta_{bc}(g_{bb} S^b_i \!+\! g_{ab}p_i S^a_i) \!+\!\sin \theta_{bc}g_{cc}S^c_i\!\Big],
\label{energy_bczeeman}\\ 
\!\!&&\!\!\mc{H}_{ab}^\text{Z}\!=\!-h\!\!\sum_i \!\Big[\!
\cos \theta_{ab}(g_{bb} S^b_i\!+\!g_{ab}p_i S^a_i)
\!+\! \sin \theta_{ab}(g_{aa} S^a_i\!+\!g_{ab}p_i S^b_i)
\!\Big], \label{energy_abzeeman}\\
\!\!&&\!\!\mc{H}_{ac}^\text{Z}\!=\!-h\!\!\sum_i \!\Big[\!
\cos \theta_{ac}(g_{aa} S^a_i\!+\!g_{ab}p_i S^b_i) \!+\! \sin \theta _{ac}g_{cc} S^c_i\!\Big],\label{energy_aczeeman}
\eea
\normalsize
where $\theta_{bc}$ and $\theta_{ab}$ denote the angle between ${\bf H}$ and ${\bf b}$, whereas $\theta_{ac}$ is the angle between ${\bf H}$ and ${\bf a}$, see insets of Fig.~\ref{fig:PhaseDiagram}. The coupling parameters are taken as in \cite{Li2019}, namely $J\!=\!0.4$~meV, $K\!=\!-18$~meV, $\Gamma\!=\!-10$~meV, $g_{aa}\!=\!g_{bb}\!=\!g_{cc}\!=\!2$ and $g_{ab}\!=\!0.1$. 

To find the classical ground states we follow the strategy developed in \cite{Rousochatzakis2018,Li2019} and use a general six-sublattice ansatz which is also cross-checked by independent unconstrained minimizations of Eq.~(\ref{eq:Hamiltonian}) on finite-size clusters. As it turns out, the classical ground state for ${\bf H}$ in the $ab$ plane can be obtained by using the ansatz reported in \cite{Rousochatzakis2018,Li2019} for ${\bf H}\!\parallel\!{\bf a}$. This is due to the fact that the ansatz for ${\bf H}\!\parallel\!{\bf b}$ is a special case of the more general ansatz for ${\bf H}\!\parallel\!{\bf a}$. The same situation takes place for ${\bf H}$ in the $bc$ plane, where the classical ground state is obtained by using  more general ansatz for ${\bf H}\!\parallel\!{\bf c}$. By contrast the classical ground states in the $ac$ plane cannot be obtained from either the ansatz for ${\bf H}\!\parallel\!{\bf a}$ or the ansatz for ${\bf H}\!\parallel\!{\bf c}$, but by a more general unconstrained six-sublattice ansatz.

\vspace*{-0.3cm}
\section{Phase diagram}\label{sec:PhaseDiagram}
\vspace*{-0.3cm}
Figure~\ref{fig:PhaseDiagram}  summarizes the emerging picture for the phase diagram as we rotate from one crystallographic axis to another. The blue solid line shows the boundary of the IC order, i.e., the critical field $H^\ast$, as obtained by tracking the associated structure factor $I_I$ (see definition in App.~\ref{app} and \cite{Li2019}), whose field evolution is shown in the insets of Fig.~\ref{fig:PhaseDiagram} for a number of representative field orientations. Altogether, the interpolation of the phase diagram between different crystallographic axes comes with a number of unexpected results. 
Most notably, we find that while the critical field $H^\ast$ grows very fast as we approach the ${\bf a}$ axis, reflecting the inequalities $H_{\bf b}^\ast\!<\!H_{\bf c}^\ast\!\ll\!H_{\bf a}^\ast$ found in \cite{Li2019}, the way $H^\ast$ grows in the $ab$-plane gives rise to a reentrant IC order in a finite range of field orientations ($\theta_{ab}\!\in\![62^\circ,69^\circ]$ and $ \theta_{ab}\!\in\![111^\circ,118^\circ]$), see inset of Fig.~\ref{fig:PhaseDiagram}\,(b). 
 In this range, the magnetic field drives the system through four different phases: the low-field six-sublattice phase (including the IC plus the uniform orders), followed by a two-sublattice order (including only the uniform orders), followed by the reentrant six-sublattice order, and then back to the two-sublattice order  (whose robust zigzag component diminishes only at $H\!\to\!\infty$).

\begin{figure}[!t]
{\includegraphics[width=\linewidth]{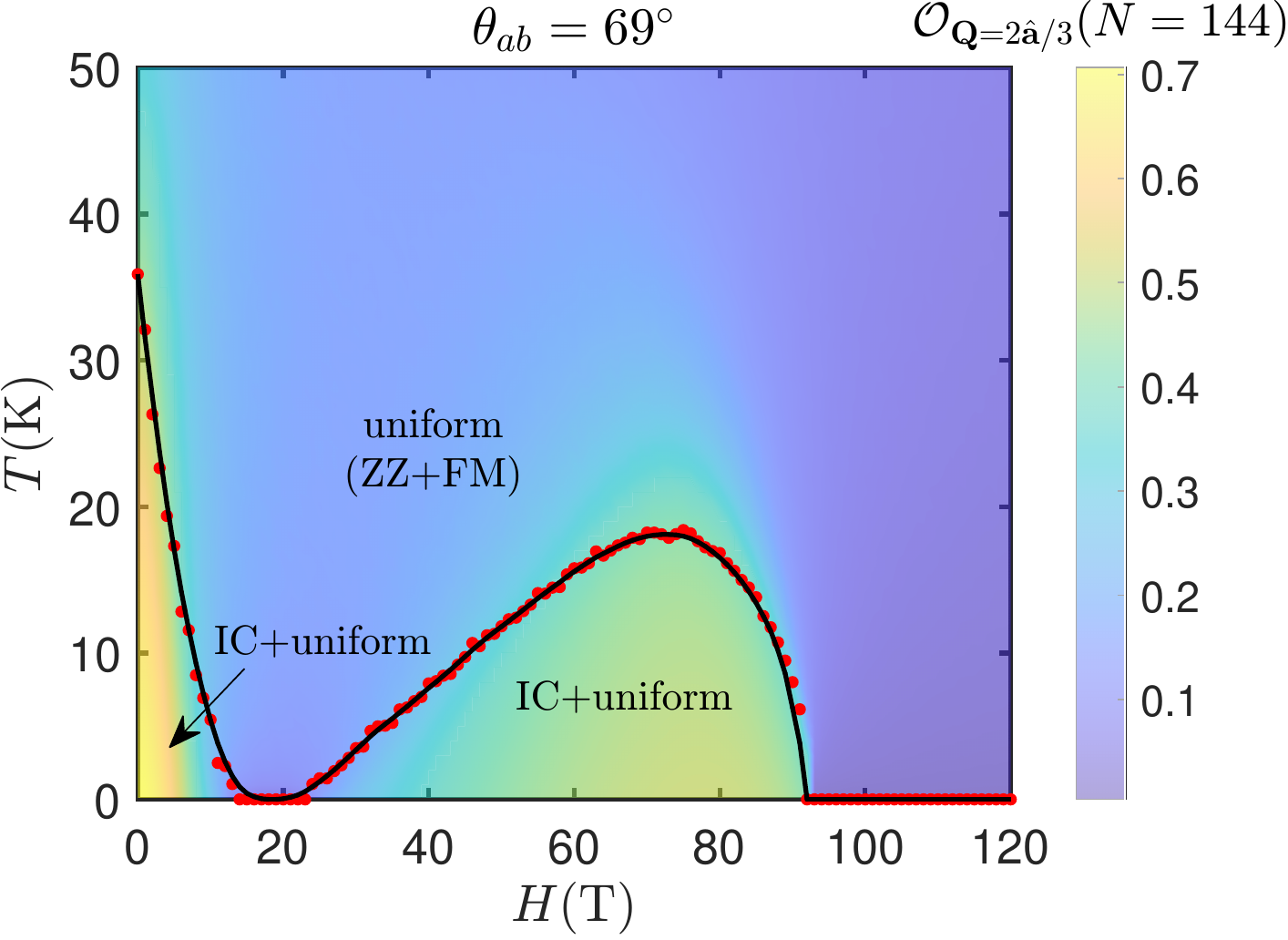}}
\caption{The field-temperature phase diagram for $H$ in the $ab$ plane with $\theta_{ab}\!=\!69^\circ$ obtained from classical MC simulations. The critical temperatures at which the modulated order disappears are shown by red dots and the black line is a guide to the eye.}\label{fig:MC}
\end{figure}

Our classical Monte Carlo (MC) simulations shown in Fig.~\ref{fig:MC} for $\theta_{ab}\!=\!69^\circ$ demonstrate that this quite unusual reentrant behavior remains robust in a finite temperature range (up to 18~K for the parameters chosen in the present study) and should therefore be observable experimentally, given in addition that the reentrance can occur at a field as low as $22$~T (for $\theta_{ab}\!=\!69^\circ$).
The critical temperatures shown by red circles in Fig.~\ref{fig:MC} are computed using the crossing of the Binder cumulant~\cite{Binder1993,Li2019} 
\be
B_{\mc{O}_{{\bf Q}=2\hat{\bf a}/3}}= 1-\langle \mc{O}^4_{{\bf Q}=2\hat{\bf a}/3}\rangle \Big{/} \Big( 3\langle \mc{O}^2_{{\bf Q}=2\hat{\bf a}/3}\rangle^2 \Big)\,,
\ee 
where $\mc{O}_{{\bf Q}=2\hat{\bf a}/3}$ is the modulated order (see App.~\ref{app} and \cite{Li2019}). 

\begin{figure}[!b]
{\includegraphics[width=0.85\linewidth]{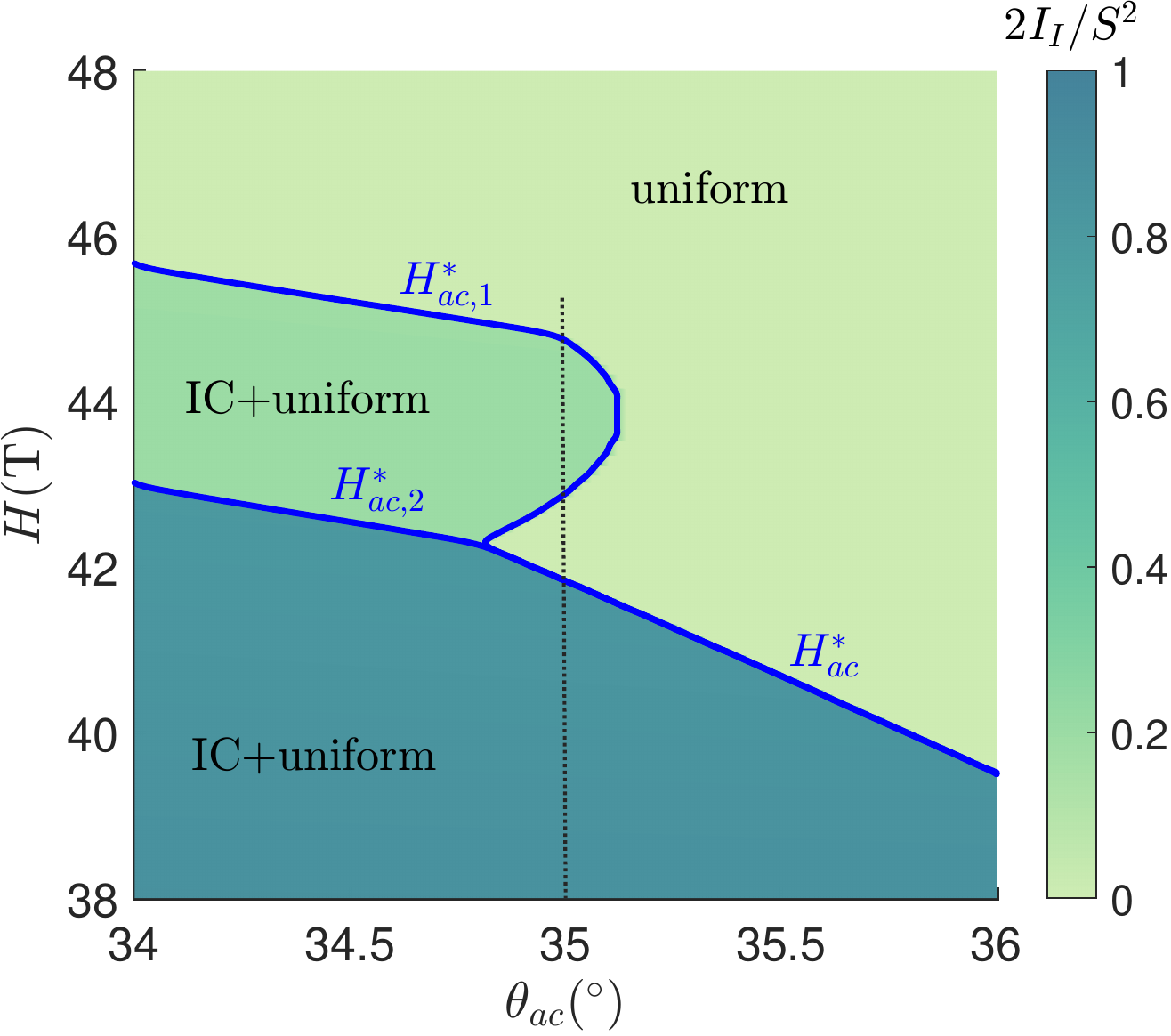}}
\caption{Evolution of the Bragg peak intensities, $I_I$, when the field direction in the $ac$-plane is within the range of angles, $\theta_{ac}\!\in\![34^\circ,36^\circ]$.}
\label{fig:ac_intensity_zoom}
\end{figure}

A similar reentrant behavior is also observed for fields in the $ac$ plane, for $\theta_{ac}\!\in\![34.8^\circ,35.1^\circ]$ (see e.g. the dashed vertical line in Fig.~\ref{fig:ac_intensity_zoom}), although this region is likely too narrow to survive at high enough temperatures to be observed. 

Figure~\ref{fig:ac_intensity_zoom} shows in addition another interesting feature, which will show up in our discussion of the torque signal below. Namely that, in the region extending for $\theta_{ac}\!\in\![17.0^\circ,34.8^\circ]$, the disappearance of the IC order proceeds with two consecutive metamagnetic jumps.  Such jumps can be seen in the inset of Fig.~\ref{fig:PhaseDiagram}\,(c). The characteristic fields at which these jumps take place are denoted by  $H_{ac,1}^*$ and $H_{ac,2}^*$ in Fig.~\ref{fig:ac_intensity_zoom}. 
The two-step disappearance of the IC order is accompanied by a similar two-step behaviour of the uniform (zigzag and ferromagnetic) orders, as seen in the evolution of the associated structure factor $I_V$ (see Fig. \ref{fig:intensity_Q0}\,(c)). This arises from the fact that the total intensity from all intertwined orders obeys the intensity sum rule $2I_I\!+\!I_V\!=\!S^2$~\cite{Rousochatzakis2018,Li2019}.
 
To conclude this section, we comment on the red dashed lines shown on top of the boundaries of the IC phase in Fig.~\ref{fig:PhaseDiagram}. These lines correspond to the following approximations for the critical fields 
\bea
bc\text{-plane:}~~&&H_{bc}^*\simeq H_{b}^*/\cos{\theta_{bc}}\,,\\
ab\text{-plane:}~~&&H_{ab}^*\simeq H_{b}^*/\cos{\theta_{ab}}\,,\\
ac\text{-plane:}~~ &&H_{ac}^*\simeq H_{c}^*/\sin{\theta_{ac}}\,.
\eea
These expressions arise from simple geometrical considerations, i.e., by associating the onset of the transition with the field at which its projection along the axes ${\bf b}$ or ${\bf c}$ reaches the critical fields $H_b^*$ or $H_c^*$, respectively. 
The expressions work surprisingly well in a a wide range of angles, especially for fields in the $bc$ and $ab$ planes. 
Importantly, the above linear relationship between $H_{bc}^*$ and $1/\cos{\theta_{bc}}$ has been reported by Modic {\it et al} for $\gamma$-$\text{Li}_2\text{IrO}_3$~\cite{Modic2014}.

\begin{figure*}
{\includegraphics[width=\textwidth]{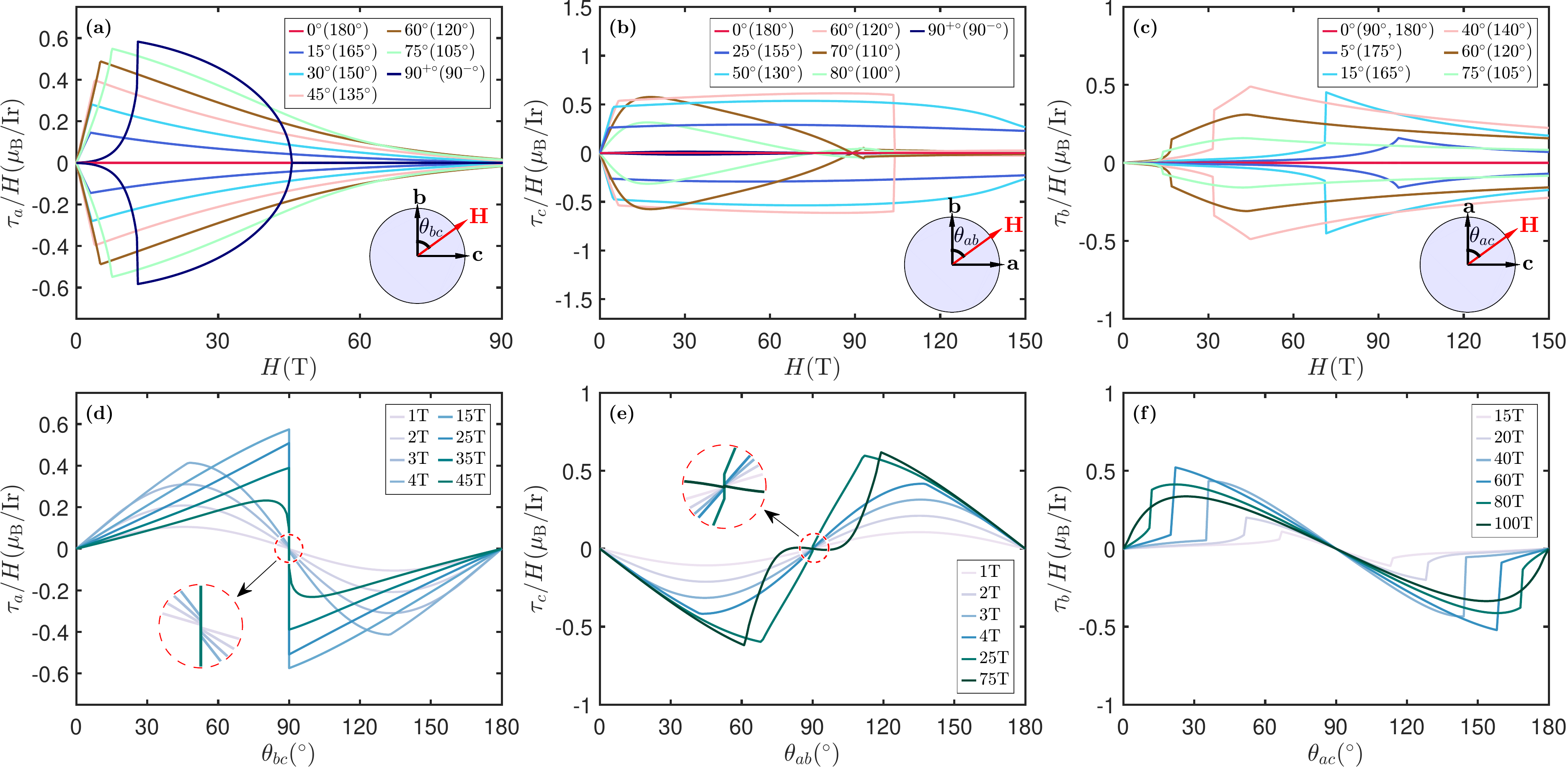}}
\caption{Field (a-c) and angular (d-f) dependence of the ratio of the torque to the magnetic field: (a,d) $\tau_a/H$ for the field applied in the $bc$ plane; (b,e) $\tau_c/H$ for the field applied in the $ab$ plane; (c,f) $\tau_b/H$ for the field applied in the $ac$ plane. The insets in (d, e) zoom-in the jumps of $\tau_a/H$ at ${\bf H}\!\parallel\!{\bf c}$ and $\tau_c/H$ at ${\bf H}\!\parallel\!{\bf a}$, respectively. These jumps originate from the explicit breaking of some of the discrete symmetries by the field component along ${\bf b}$~\cite{Li2019}.}\label{fig:torque}
\end{figure*}

\vspace*{-0.3cm}
\section{Magnetic torque}\label{sec:torque_abc}
\vspace*{-0.3cm}
Having established the general form of the phase diagram we now turn to the magnetic torque, whose behavior as a function of field strength and orientation is summarized in Fig.~\ref{fig:torque}. 
The torque ${\boldsymbol \tau}\!=\!(\tau_a,\tau_b,\tau_c)$ is given as ${\boldsymbol \tau}\!=\!{\bf m}\!\times\!{\bf H}$, where ${\bf m}\!=\!(m_a,m_b,m_c)$ is the magnetization per site.
The data shown in Fig.~\ref{fig:torque} reveal a wealth of complex features that are distinctive for each crystallographic plane. Some of these features reflect naturally the phase transitions discussed above, but there are again some additional striking features of experimental relevance. 
Among these, the most notable is the large discontinuous reversal of the torque at ${\bf H}\parallel{\bf c}$ as the field rotates in the $bc$ plane, and its characteristic `sawtooth' shape (see Fig.~\ref{fig:torque}\,(d)) that resembles closely the one reported in $\gamma$-Li$_2$IrO$_3$~\cite{Modic2014,Modic2017}. 
This discontinuity was already predicted in \cite{Li2019} and is attributed to the fact that the uniform (zigzag and ferromagnetic) orders that drive the torque are chosen spontaneously inside the region $0\!<\!H\!\le\!H_c^{**}$ for ${\bf H}\!\parallel\!{\bf c}$.
A torque reversal of similar origin occurs at ${\bf H}\!\parallel\!{\bf a}$ when ${\bf H}$ rotates in the $ab$ plane (and for $0\!<\!H\!\le\!{ H_a^\ast}$), but the height of this reversal is likely too small to be observed.   

Before we discuss these features in detail let us first briefly recall the main aspects of the torques for ${\bf H}\!\parallel\!{\bf a}$, ${\bf H}\!\parallel\!{\bf b}$ and ${\bf H}\!\parallel\!{\bf c}$~\cite{Li2019}, which are included in the insets of Fig.~\ref{fig:torque}.

\vspace*{-0.3cm}
\subsection{Magnetic torque for fields along the crystallographic axes}
\vspace*{-0.3cm}
The torque for ${\bf H}\!\parallel\!{\bf b}$ is zero for all the fields. For ${\bf H}\!\parallel\!{\bf a}$ and ${\bf H}\!\parallel\!{\bf c}$, the torques are, correspondingly, along $\bf c$ and $\bf a$. The torque for ${\bf H}\!\parallel\!{\bf a}$ remains non-zero up to the critical field $H_{\bf a}^{\ast}$, whereas the torque for ${\bf H}\!\parallel\!{\bf c}$ remains non-zero up to the second critical field $H_{\bf c}^{\ast\ast}$, which marks the transition into the fully polarized state~\cite{Li2019}. 
In both cases, the sign of the torque is chosen spontaneously and its magnitude is proportional to the $m_b$-component of the magnetization, $\tau_c/H\!=\!m_b$ and $\tau_a/H\!=\!m_b$. Note, however, that $m_b$ is significantly different in the two cases, leading to a much larger $|\tau_a|$ for fields along ${\bf c}$ compared to $|\tau_c|$ for fields along ${\bf a}$~\cite{Li2019}. 
Note also that for ${\bf H}\!\parallel\!{\bf c}$, the zigzag and ferromagnetic orders are comparable to each other, but the dominant contribution to $\tau_a$ comes from the ferromagnetic order, since $g_{bb}\gg g_{ab}$. 
Moreover, the uniform orders have a non-monotonic behavior as a function of $H$ and exhibit a finite jump at the critical field $H_{\bf c}^\ast$, leading to a sharp kink in $\tau_a$, see dark blue line in Fig.~\ref{fig:torque}\,(a) associated with $\theta_{bc}\!=\!90^\circ$.

\vspace*{-0.3cm}
\subsection{Magnetic torque for fields away from the crystallographic axes}\label{sec:torque_general}
\vspace*{-0.3cm}
{\it Fields within the $bc$ plane.} 
This is the setup explored experimentally in $\gamma$-$\text{Li}_2\text{IrO}_3$~\cite{Modic2014,Modic2017}. Here the torque is along the ${\bf a}$ axis, 
with 
\be\label{eq:taua}
\tau_a/H=m_b \sin\theta_{bc}-m_c\cos\theta_{bc}\,.
\ee
The sign of $\tau_a$ for ${\bf H}\!\parallel\!{\bf c}$ ($\theta_{bc}\!=\!90^\circ$) is chosen spontaneously below $H_{\bf c}^{\ast\ast}$, is zero for ${\bf H}\!\parallel\!{\bf b}$ ($\theta_{bc}\!=\!0^\circ$)~\cite{Li2019} and is chosen by the field for any intermediate angles.

The corresponding torque data are shown in Figs.~\ref{fig:torque}\,(a, d). At low $\theta_{bc}$ and $H$ we find $\tau_a\!\propto\!H^2\sin\theta_{bc}$, which is the expected behavior in the linear magnetization regime, where $m_b\!=\!\chi_b H\cos{\theta_{bc}}$, $m_c\!=\!\chi_c H\sin{\theta_{bc}}$ and $\tau_a\!=\!\frac{1}{2}(\chi_b\!-\!\chi_c)H^2\sin{2\theta_{bc}}$, where $\chi_b$ and $\chi_c$ are the magnetic susceptibilities.
This behavior ends with a kink at the critical field $H^*_{bc}$ associated with the disappearance of the IC order, which happens via a second order phase transition. 

With increasing $\theta_{bc}$, the low-field behavior of $\tau_a/H$ turns from linear to nonlinear (see Fig.~\ref{fig:torque}\,(a)), and for $\theta_{bc}$ close to $90^\circ$ the kink at $H^*$ turns to a discontinuity, consistent with previous results for a metamagnetic transition at ${\bf H}\!\parallel\!{\bf c}$~\cite{Li2019}. 
The size of the discontinuity depends non-monotonously on $\theta_{bc}$, with the largest jump taking place at $\theta_{bc}\!=\!90^\circ$.  
At this angle, there is a second phase transition at $H_{\bf c}^{**}$, above which the classical magnetization saturates and the torque vanishes identically (dark blue line in Fig.~\ref{fig:torque} (a)). 
Recall that the presence of the second transition at high field is due to the fact that the zigzag order for ${\bf H}\!\parallel\!{\bf c}$ can take two possible orientations, and the system chooses one of them spontaneously. For all other fields in the $bc$-plane, the orientation of the zigzag order is determined by the field and the second phase transition is absent. For all $\theta_{bc}\!\neq\!90^\circ$, the torque is slowly approaching zero with increasing field, and becomes identically zero only at infinite field.

\begin{figure}[!b]
{\includegraphics[width=0.8\linewidth]{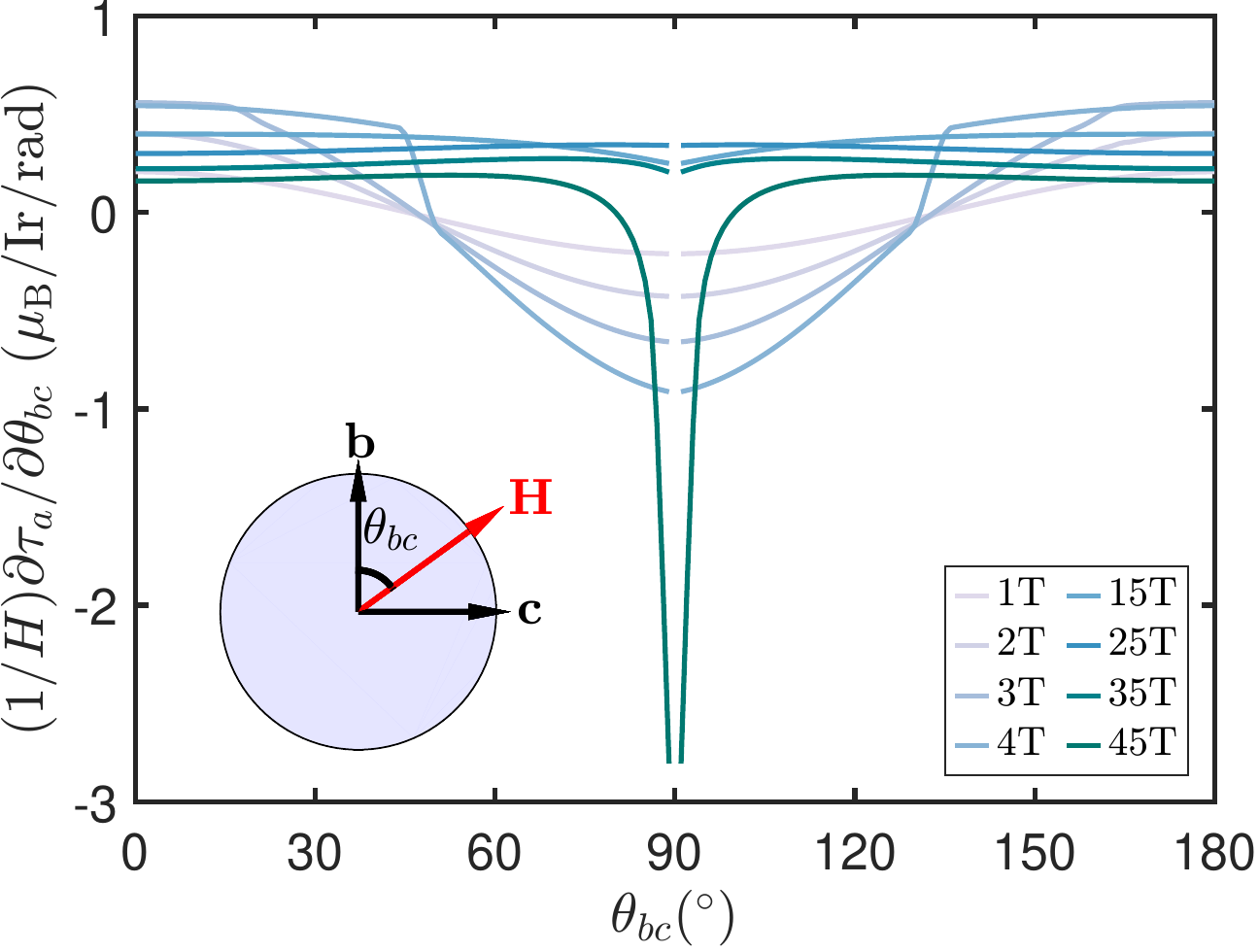}}
\caption{Magnetotropic coefficient $k\!=\!(1/H)\partial\tau_a/\partial\theta_{bc}$ as a function of $\theta_{bc}$ for several field strengths. The magnetotropic coefficient at $\theta_{bc}\!=\!90^\circ$ is not well-defined due to the discontinuity of $\tau_a$.} \label{fig:bc_torque_derivative}
\end{figure}

Turning to the angular dependence of the torque (see Fig.~\ref{fig:torque} (d)), the low-field sinusoidal behavior turns into a characteristic sawtooth shape at high fields, featuring a large torque reversal at $\theta_{bc}\!=\!90^\circ$. Such a reversal is also present at lower fields, but becomes much more pronounced at higher fields. The reversal is due to the explicit breaking of the discrete symmetries $\Theta C_{2{\bf b}}$ and $C_{2{\bf c}}$ by the component of the field along ${\bf b}$~\cite{Li2019}. 
Note that the sawtooth torque profile will give rise to a pronounced anomaly in the magnetotropic coefficient $k\!=\!(1/H)\partial\tau_a/\partial\theta_{bc}$ (see Fig.~\ref{fig:bc_torque_derivative}). This thermodynamic quantity characterizes the curvature of the free energy with respect to the field orientation, and can be measured directly via torsion magnetometry~\cite{Modic2018a}.

{\it Field within the $ab$ plane.} 
Because of a very strong anisotropy of the critical fields, $H_a^*/H_b^*\!\simeq\!50$, a rather different picture is observed for ${\bf H}$ in the $ab$-plane. In this setup the torque is along the ${\bf c}$ axis,  with
\be\label{eq:tauc}
\tau_c/H=m_a \cos\theta_{ab}-m_b \sin\theta_{ab}\,.
\ee
The corresponding data are shown in Fig.~\ref{fig:torque}\,(b, e). Here the low-field behavior $\tau_c\!\propto\!H^2\sin\theta_{ab}$ persists up to a critical field $H^*_{ab}\!\simeq\!H^*_{b}$. At this field the torque shows a kink which persists for a large range of angles, $\theta_{ab}\!\leqslant\!50^\circ$ and $\theta_{ab}\!\geqslant\!130^\circ$. This behavior is similar to the case of the field sweeping in the $bc$ plane except that at $H\!>\!H^*_{ab}$ the torque shows almost field-independent behavior up to high fields, $H\!\sim\! H^*_{a}$, and then slowly decreases. This behavior reflects the robustness of the uniform orders contributing to $m_a$ and $m_b$.
  
As the field gets closer to ${\bf a}$, the contribution from the ${\bf a}$-component of the field increases. For the ranges $\theta_{ab}\!\in\![50^\circ,62^\circ]$ and $\theta_{ab}\!\in\![118^\circ,130^\circ]$, $\tau_c/H$ then exhibits two sharp kinks appearing due to the competition between the two contributions in the Zeeman term in Eq.~(\ref{energy_abzeeman}). In particular, Fig.~\ref{fig:torque}\,(b) shows that  $\tau_c/H$ for $\theta_{ab}=60^\circ(120^\circ)$ has two  kinks, one at low field close to $H_b^*$ and the other at intermediate field close to $H_a^*$.  

Especially interesting is the behavior of the torque for $\theta_{ab}\!\in\![62^\circ,69^\circ]$ and $\theta_{ab}\!\in\![111^\circ,118^\circ]$. Here, the competition between the two contributions in Eq.~(\ref{energy_abzeeman}) leads to the unusual reappearance of the modulated order at intermediate fields, as discussed above. As the field further approaches the $\bf a$-axis ($70^\circ\!\leqslant\!\theta_{ab}\leqslant\!110^\circ$), these two kinks disappear but another interesting feature is observed. At the angles and magnitudes of the field at which the first and the second term in Eq.~(\ref{eq:tauc}) become equal, the magnetic torque simply vanishes. In Fig.~\ref{fig:torque}\,(e) this situation is shown for $\theta_{ab}\!=\!70^\circ(110^\circ)$ and $\theta_{ab}\!=\!80^\circ(100^\circ)$. Finally, when $\theta_{ab}\!=\!90^\circ$ (${\bf H}\!\parallel\!{\bf a}$), the torque signal is barely visible since the contribution of the first term in Eq.~(\ref{eq:tauc}) becomes vanishingly small for all values of $H$. Also, for ${\bf H}\!\parallel\!{\bf a}$ the sign of the ferromagnetic uniform order ($\parallel{\bf b}$) is spontaneously chosen, which results in a small jump in the torque as shown in the inset of Fig.~\ref{fig:torque}(e). 

\begin{figure}
{\includegraphics[width=0.95\linewidth]{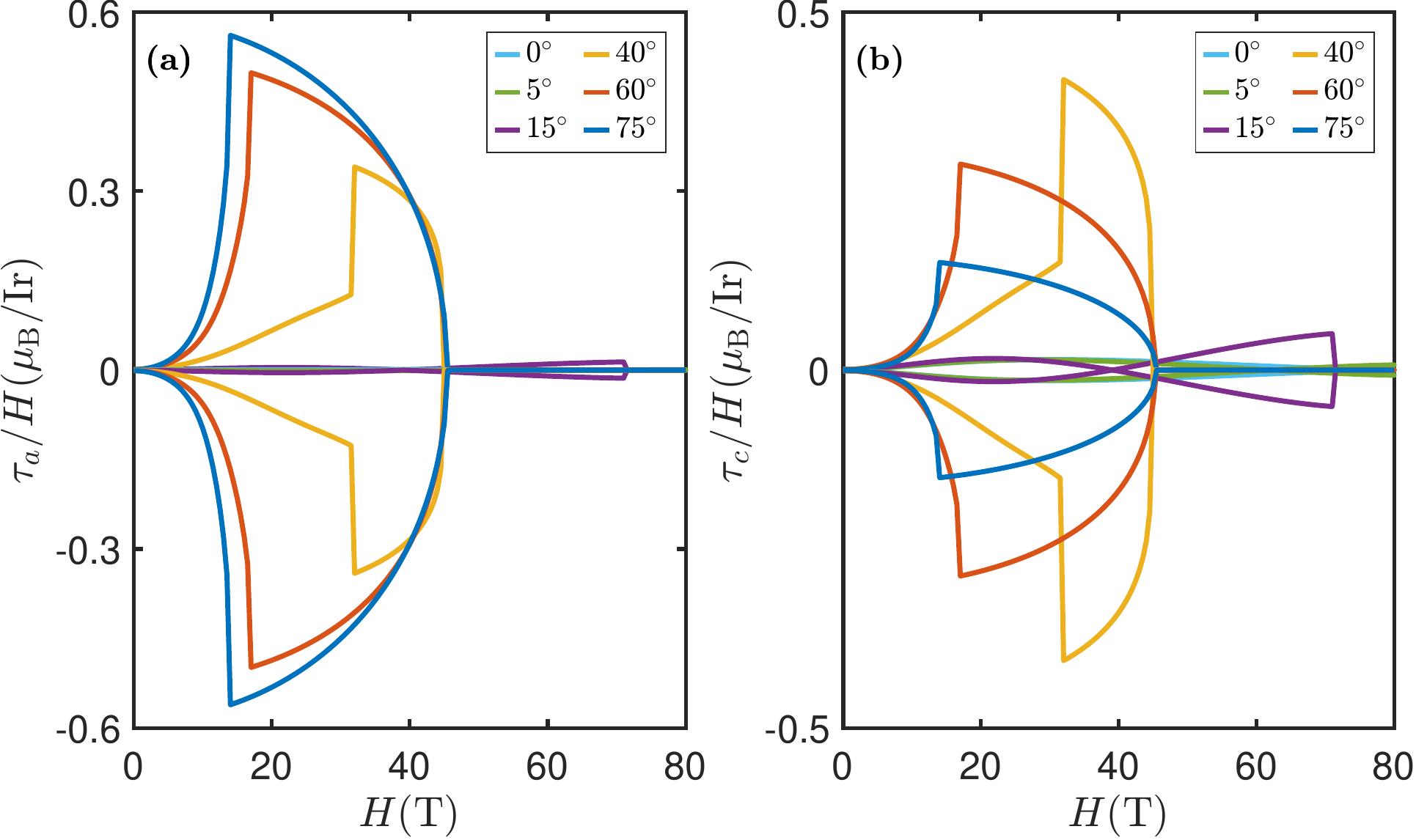}}
\caption{Evolution of (a) $\tau_a/H$ and (b) $\tau_c/H$ as a function of $H$ for fields rotating in the $ac$ plane, for several field orientations $\theta_{ac}$. To each angle there correspond two curves with opposite torque signs, reflecting the fact that this sign is chosen spontaneously when ${\bf H}\perp{\bf b}$.}\label{fig:ac_torque_2}
\end{figure}

\begin{figure}
\includegraphics[width=0.49\textwidth]{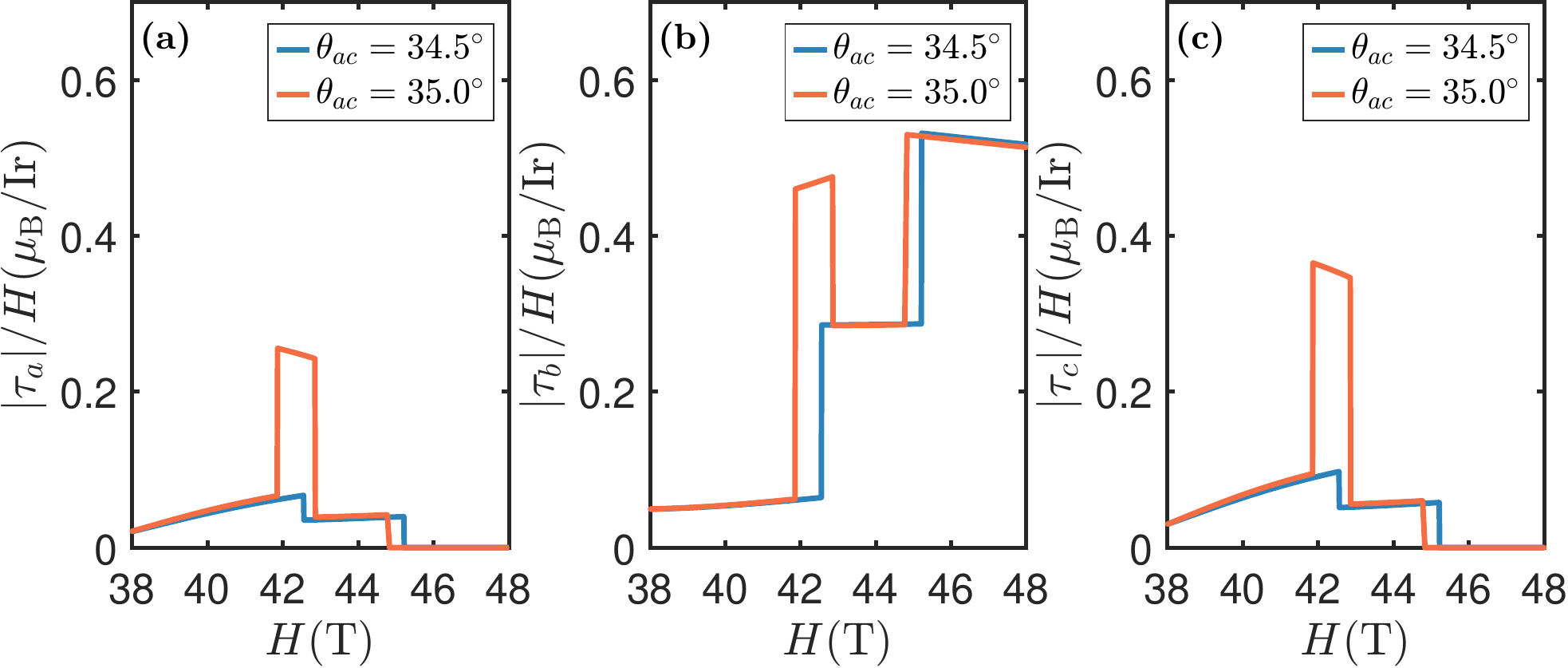}
\caption{Non-monotonous evolution of $|\tau_a|/H$ (a),  $|\tau_b|/H$ (b) and $|\tau_c|/H$ (c) as  a function of $H$ at $\theta_{ac}\!=\!34.5^\circ$ and $\theta_{ac}\!=\!35^\circ$, where the destruction of the IC order happens via two-step process. At $\theta_{ac}\!=\!35^\circ$ the reentrant behavior is observed in addition to the two-step process.}\label{fig:actorquereentrant}
\end{figure}

{\it Field within the $ac$ plane.} 
Unlike the two previous cases, when we sweep the field in the $ac$ plane the torque is no longer perpendicular to the plane of the field. In particular, all three components $\tau_a$, $\tau_b$ and $\tau_c$  remain finite for all angles $\theta_{ac}\!\notin\!\{0^\circ,90^\circ,180^\circ\}$. This aspect leads to a complex non-monotonic evolution of the torque. Here, the torque components are related to the magnetization components via
\begin{align}\label{eq:tau}    
\tau_a/H\!&=\!m_b \sin\theta_{ac},\nn\\
\tau_b/H\!&=\!m_c\cos\theta_{ac}\!-\!m_a \sin\theta_{ac},\\
\tau_c/H\!&=\!-m_b \cos\theta_{ac}.\nn
\end{align}
Our numerical data for $\tau_b$ are shown in Figs.~\ref{fig:torque}\,(c, f), while those for $\tau_a$ and $\tau_c$ are shown in Figs.~\ref{fig:ac_torque_2} \,(a, b). Note that while the sign of $\tau_b$ is fixed by the field, the signs of $\tau_a$ and $\tau_c$ are chosen spontaneously for all values of $\theta_{ac}$. These two components are proportional to the magnetization along the $\bf b$-axis, $m_b$. Since the sign of $m_b$ is spontaneously chosen, the signs of $\tau_a$ and $\tau_c$ are spontaneously chosen as well. The two choices of signs are shown by the two sets of branches in Fig.~\ref{fig:ac_torque_2} associated with each $\theta_{ac}$.

As we can see from Figs.~\ref{fig:torque}\,(c) and \ref{fig:ac_torque_2}, at low fields all three components of the torque are small. At intermediate fields, the overall magnitude of the torque is growing and the three components exhibit a strong angular dependence. At small angles $\theta_{ac}$, $\tau_b/H$ first grows very slowly with the field and then exhibits a kink at $H_{ac}^{*}$ where $H_c^{**}\!<\!H_{ac}^{*}\!\lesssim \!H_a^{*}$ (see, e.g., blue line for $\theta_{ac}\!=\!5^\circ(175^\circ)$ in Fig.~\ref{fig:torque}\,(c)).  At intermediate and large angles, at which the field direction is  relatively close to the $\bf c$-axis,  $\tau_b/H$  exhibits a jump at $H_{ac}^{*}\!\gtrsim\! H_c^{*}$, indicating a first order phase transition, and then exhibits a kink at $H_{ac}^{**}\!\lesssim\! H_a^{*}$ (see, e.g., rose line for $\theta_{ac}\!=\!40^\circ(140^\circ)$ in Fig.~\ref{fig:torque}\,(c)). The angular behavior of $\tau_b/H$ shown in  Fig.~\ref{fig:torque}\,(f) for various field strengths also reflects this behavior.
 
Two critical fields at some field directions are also clearly seen in $\tau_a/H$ and $\tau_c/H$ (see Fig.~\ref{fig:ac_torque_2}). At angles close to the $\bf c$-axis, both $\tau_a/H$ and $\tau_c/H$ show a jump at  $H_{ac}^*\!\gtrsim\!H_c^*$ and then go to zero above the critical field $H_{ac}^{**}\!\simeq\!H_c^{**}$. When the orientation of the field is closer to the $\bf a$-axis, both $\tau_a/H$ and $\tau_c/H$ remain negligibly small, and the total torque is oriented almost  along the $\bf b$-axis.

Finally, let us discuss what happens with the torque in the region depicted in Fig.~\ref{fig:ac_intensity_zoom}, which includes a narrow range (around $\theta_{ac}\!=\!35^\circ$) featuring a reentrant IC phase and a more extended range ($\theta_{ac}\!\in\![17.0^\circ,34.8^\circ]$) featuring a two-step disappearance of the IC order. 
Figure~\ref{fig:actorquereentrant} shows what happens at two representative angles, $\theta_{ac}\!=\!35^\circ$ and $\theta_{ac}\!=\!34.5^\circ$, one from each range. 
As expected, the torque components show three abrupt discontinuities at $\theta_{ac}\!=\!35^\circ$ and two discontinuities at  $\theta_{ac}\!=\!34.5^\circ$, reflecting the respective evolution of the IC order.

\begin{figure*}
	{\includegraphics[width=\textwidth]{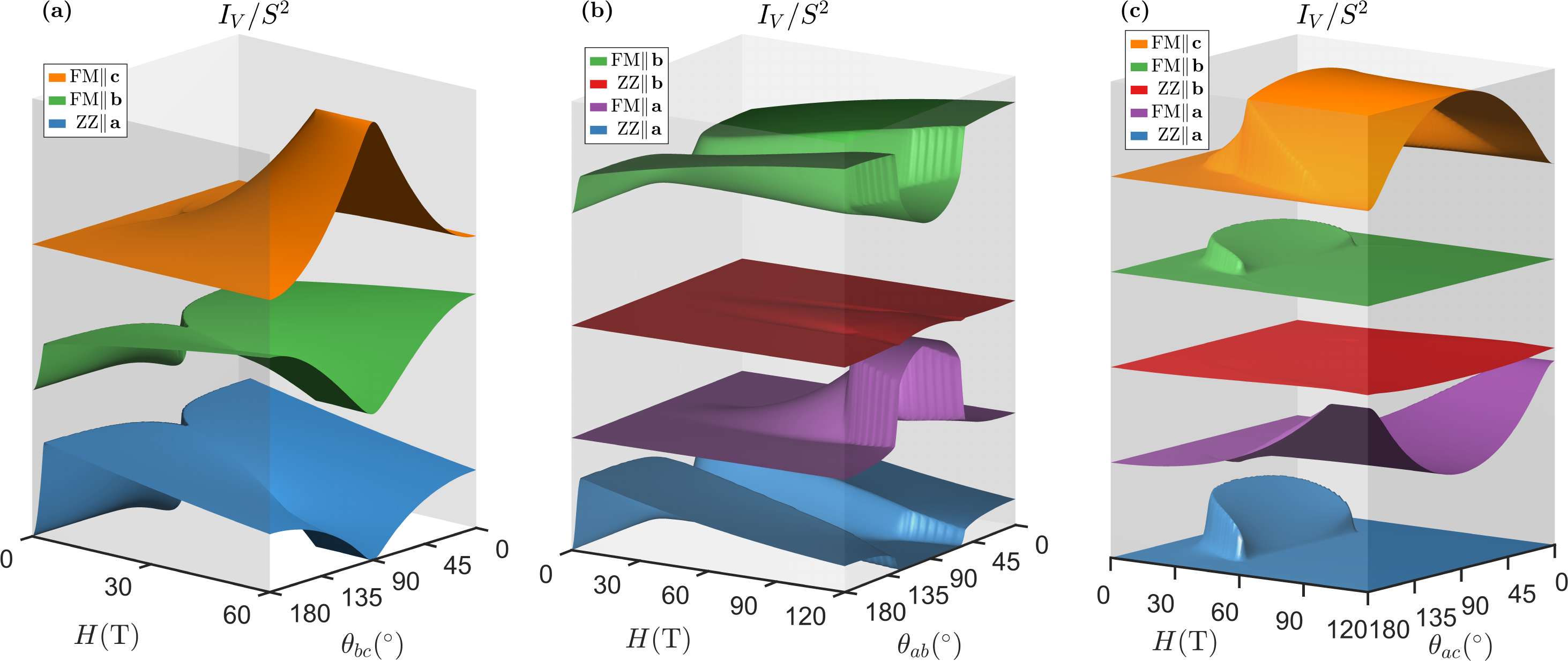}}
	\caption{Evolution of  the individual Bragg peak intensities  from the uniform  orders with field in (a) $bc$-plane, (b) $ab$-plane, and (c) $ac$-plane.}
	\label{fig:intensity_Q0}
\end{figure*}

\section{Discussion}
The above numerical results demonstrate a rather complex interplay between the modulated and uniform (zigzag and ferromagnetic) orders of $\beta$-$\text{Li}_2\text{IrO}_3$ as we rotate the field in different crystallographic planes. The resulting phase diagram shows a number of unexpected results. 
Most saliently, we find a rare example of a reentrant IC phase for fields rotating in the $ab$ plane. The stability region of this reentrant phase extends in a finite range of field orientations ($\theta_{ab}\!\in\![62^\circ,69^\circ]$ and $\theta_{ab}\!\in\![111^\circ,118^\circ]$), field strengths and temperatures. This unusual behavior is yet another manifestation of the strongly anisotropic character of the underlying microscopic interactions in $\beta$-$\text{Li}_2\text{IrO}_3$, and we have provided numerical estimates for its experimental verification. 
A similar reentrant behavior is also observed for fields sweeping in the $ac$-plane but only for a very narrow range of angles.

Our numerical results for the behavior of the magnetic torque offers further insights and reveals a number of distinctive signatures that are unique for each crystallographic plane. 
Quite generally, the sequence of field-induced phase transitions give rise to torque anomalies (jumps and kinks) which can be directly seen with torque magnetometry. 
Among the distinctive aspects of the torque is the observation of a large discontinuous reversal as the field rotates in the $bc$ plane and passes through the ${\bf c}$ axis. The field orientation profile of this sharp discontinuity resembles closely the sawtooth-like behaviour reported by Modic {\it et al} in $\gamma$-$\text{Li}_2\text{IrO}_3$~\cite{Modic2014,Modic2017}. 
In that compound the discontinuity extends well beyond $H_c^*\!\simeq\!16$~T and persists up to 55~T, which likely corresponds to the specific value of $H_c^{**}$ in $\gamma$-$\text{Li}_2\text{IrO}_3$~\cite{Modic2017}. While Modic {\it et al}~\cite{Modic2018b} attributed this torque anomaly of $\gamma$-$\text{Li}_2\text{IrO}_3$ to the presence of a chiral spin-order, the more conventional  symmetry breaking mechanism presented here for $\beta$-$\text{Li}_2\text{IrO}_3$ appears more likely, given the close similarity of the two compounds. 

The experimental observation of the above distinctive features can be used to map out a quantitative phase diagram and extract accurate values of the microscopic exchange couplings of $\beta$-$\text{Li}_2\text{IrO}_3$. This compound is rather unique in this respect, because its microscopic model has been better understood~\cite{Lee2015,Lee2016,Ducatman2018,Rousochatzakis2018,Li2019} compared to its 2D and 3D analogues, $\alpha$- and $\gamma$-$\text{Li}_2\text{IrO}_3$.

\vspace*{0.3cm} 
\noindent{\it  Acknowledgments:} 
We thank J. Analytis, J. Betouras,  K. Modic,  A. Ruiz and A. A. Tsirlin  for helpful discussions. This work was supported by the U.S. Department of Energy, Office of Science, Basic Energy Sciences under Award No. DE-SC0018056. We also acknowledge the support of the Minnesota Supercomputing Institute (MSI) at the University of Minnesota.

\appendix
\titleformat{\section}{\normalfont\bfseries\filcenter}{Appendix~\thesection:}{0.25em}{}

\vspace*{-0.3cm}

\section{Static structure factors and Bragg peak intensities}\label{app}
In this appendix, we briefly review our definitions of the symmetry-resolved static structure factors.
For the various components of the static structure factor, we follow the convention of Ref.~\cite{Ducatman2018} and denote the modulated components with ${\bf Q}\!=\!2\hat{{\bf a}}/3$ by the letter $M$ and the uniform components with ${\bf Q}\!=\!0$ by $M'$. 
Specifically, the modulated ${\bf Q}\!=\!2\hat{\bf a}/3$ components read
\begin{align}
\begin{pmatrix}
	i\,{\bf M}(A)\\
	i\,{\bf M}(C)\\
	{\bf M}(F)\\
	i\,{\bf M}(G)\\
\end{pmatrix} \!\equiv\! \frac{1}{4}
\begin{pmatrix}
	{\bf S}_1({\bf Q})-{\bf S}_2({\bf Q})-{\bf S}_3({\bf Q})+{\bf S}_4({\bf Q})\\
	{\bf S}_1({\bf Q})+{\bf S}_2({\bf Q})-{\bf S}_3({\bf Q})-{\bf S}_4({\bf Q})\\
	{\bf S}_1({\bf Q})+{\bf S}_2({\bf Q})+{\bf S}_3({\bf Q})+{\bf S}_4({\bf Q})\\
	{\bf S}_1({\bf Q})-{\bf S}_2({\bf Q})+{\bf S}_3({\bf Q})-{\bf S}_4({\bf Q})
\end{pmatrix}_{{\bf Q}=2\hat{\bf a}/3}\!\!\!\!,
\end{align}
and the uniform ${\bf Q}\!=0$ components are defined as
\begin{align}
\begin{pmatrix}
	{\bf M}'(A)\\
	{\bf M}'(C)\\
	{\bf M}'(F)\\
	{\bf M}'(G)\\
\end{pmatrix} \equiv \frac{1}{4}
\begin{pmatrix}
	{\bf S}_1({\bf 0})-{\bf S}_2({\bf 0})-{\bf S}_3({\bf 0})+{\bf S}_4({\bf 0})\\
	{\bf S}_1({\bf 0})+{\bf S}_2({\bf 0})-{\bf S}_3({\bf 0})-{\bf S}_4({\bf 0})\\
	{\bf S}_1({\bf 0})+{\bf S}_2({\bf 0})+{\bf S}_3({\bf 0})+{\bf S}_4({\bf 0})\\
	{\bf S}_1({\bf 0})-{\bf S}_2({\bf 0})+{\bf S}_3({\bf 0})-{\bf S}_4({\bf 0})
\end{pmatrix} \,,
\end{align}
where ${\bf S}_{\nu=1-4}$ represent the four spin sites in each primitive unit cell. The relative amplitudes of ${\bf S}_{\nu}$ are characterized by the four-component symmetry basis vectors 
\begin{align}\label{eq:symmetrybasis}
A\!=\!\begin{pmatrix} 1\\-1\\-1\\1\end{pmatrix},~~~ 
C\!=\!\begin{pmatrix} 1\\1\\-1\\-1\end{pmatrix},~~~  
F\!=\!\begin{pmatrix}1\\1\\1\\1\end{pmatrix},~~~  
G\!=\!\begin{pmatrix}1\\-1\\1\\-1\end{pmatrix}\,,  
\end{align}
where $A$, $C$, $F$, and $G$ denote the N\'eel, stripy, ferromagnetic, and zigzag order, respectively~\cite{Biffin2014a}.

Given the static structure factor components, we can compute the corresponding contribution to the Bragg peak intensities $I_{\text{tot}}=2I_I+I_V$. We first examine the modulated order contribution which is given by
\begin{align}
I_I=\sum_{i=1}^4 I_{I,~\Gamma_i}\simeq I_{I,\Gamma_4} = |M_a(A)|^2+|M_b(C)|^2+|M_c(F)|^2 ,
\end{align}
where $\Gamma_{i=1-4}$ are the irreducible representations associated with the basis vectors in Eq.~(\ref{eq:symmetrybasis}). It turns out that $\Gamma_4 = (A,C,F)$ always provides the dominant contribution~\cite{Li2019}. The evolution of $I_I$ for field within each  crystallographic plane is given in Fig.~\ref{fig:PhaseDiagram} of the main text.

The uniform orders can have six different components, which correspond to the zigzag (ZZ) order or ferromagnetic (FM) order along $\bf a$, $\bf b$, or $\bf c$ direction, i.e., $M_a'(G)$, $M_a'(F)$, $M_b'(G)$, $M_b'(F)$, $M_c'(G)$, and $M_c'(F)$. As discussed in Ref.~[\onlinecite{Li2019}], when the field is applied along one of the orthorhombic axes, the  uniform components are:
\begin{align}\label{eq:UniformOrder}
{\bf H\parallel a}:  &~ M_a'(G), M_a'(F), M_b'(G), M_b'(F), \nn\\
{\bf H\parallel b}:  &~ M_a'(G), M_b'(F), \nn\\
{\bf H\parallel c}:  &~ M_a'(G), M_b'(F), M_c'(F) .
\end{align}
Note that for ${\bf H\parallel c}$, $M_a'(G)$ and $M_b'(F)$ only remain finite for $H\!<\!H_c^{**}$. Correspondingly, when the field is rotated on  the  crystallographic planes, there is always a competition between those components in Eq.~(\ref{eq:UniformOrder}). Figure~\ref{fig:intensity_Q0} shows the individual non-vanishing contributions $I_{V,i}=|M_i'|^2$ of each uniform component to the total Bragg peak intensity $I_V=\sum_{i=1}^6 I_{V,i}$. This complicated evolution clearly signifies the anisotropic character of the Hamiltonian  in Eq.~(\ref{eq:Hamiltonian}).

\bibliography{reference}

\end{document}